\RequirePackage{fix-cm}

\documentclass[twocolumn]{svjour3}
\smartqed  
\usepackage{graphicx,amsmath,xcolor,subcaption,natbib,amssymb}
\usepackage[labelformat=simple]{subcaption}
\usepackage{comment,soul}
\usepackage{color,hyperref}
\hypersetup{colorlinks=true,
	linkcolor=blue,
	citecolor=blue}

\begin{document}
	
	\title{Spectral Analysis Modal Methods (SAMMs) using Non-Time-Resolved PIV
	}
	
	
	\author{YANG ZHANG         \and
		LOUIS N. CATTAFESTA III \and LAWRENCE UKEILEY 
	}
	
	
	\institute{YANG ZHANG, LOUIS N. CATTAFESTA III \at
		Florida Center for Advanced Aero‑Propulsion (FCAAP),
		FAMU‑FSU College of Engineering, Florida State
		University, 2003 Levy Ave, Tallahassee, FL 32310, USA
		\\
		\email{lcattafesta@fsu.edu}           
		\and
		LAWRENCE UKEILEY \at
		University of Florida, Gainesville, Florida, 32611
	}
	
	\date{Received: date / Accepted: date}

	\maketitle
	
	\begin{abstract}
		
		We present spectral analysis modal methods (SAMMs) to perform POD in the frequency domain using non-time-resolved Particle Image Velocity (PIV) data combined with unsteady surface pressure measurements. In particular, time-resolved unsteady surface pressure measurements are synchronized with non-time-resolved planar PIV measurements acquired at 15 Hz in a Mach 0.6 cavity flow. Leveraging the spectral linear stochastic estimation (LSE) method of \citet{Tinney2006}, we first estimate the cross correlations between the velocity field and the unsteady pressure sensors via sequential time shifts, followed by a Fast Fourier transform to obtain the pressure-velocity cross spectral density matrix. This leads to a linear multiple-input / multiple-output (MIMO) model that determines the optimal transfer functions between the input cavity wall pressure and the output velocity field. Two variants of SAMMs are developed and applied. The first, termed ``SAMM-SPOD'', combines the MIMO model with the SPOD algorithm of \citet{Towne2018}.  The second, called ``SAMM-RR'', adds independent sources and uses a sorted eigendecomposition of the input pressure cross-spectral matrix to enable an efficient reduced-rank eigendecomposition of the velocity cross-spectral matrix. In both cases, the resulting rank-1 POD eigenvalues associated with the Rossiter frequencies exhibit very good agreement with those obtained using independent time-resolved PIV measurements. The results demonstrate that SAMMs provide a methodology to perform space-time POD without requiring a high-speed PIV system, while avoiding potential pitfalls associated with traditional time-domain LSE. \end{abstract}
	

	\section{Introduction}
	\label{intro}
	Modal methods provide powerful tools for fluid dynamics, encompassing Proper Orthogonal Decomposition (POD) \citep{Lumley1967}, dynamic mode decomposition \citep{Schmid2010a}, and resolvent analysis \citep{Mckeon2010}. They provide fundamental understanding of turbulent flows, for example to identify coherent structures \citep{Berkooz1993}, develop reduced-order models \citep{Pinnau2008}, or facilitate flow control. These popular methods are recently reviewed by \citet{Taira_2017,Taira:2019bk}. Applications of the POD are associated with the solution of an eigenvalue problem which yields an orthogonal set of basis functions (eigenvectors) whose importance are represented by the value of their corresponding eigenvalue.  Snapshot POD uses instantaneous independent snapshots of the flow at different time instances to decompose a flow field into spatially orthonormal modes \citep{Sirovich1987}. The velocity field at the times corresponding to those of the snapshots can then be reconstructed using a linear combination of time-dependent POD coefficients and the corresponding spatial modes. Note that this popular version of POD produces spatially coherent structures. Time-resolved (TR) data acquisition is generally required to obtain unambiguous dynamical information. While feasible in unsteady simulations, these tools remain elusive in experiments due to the high cost or limitations of current instrumentation. 
	
	The velocity field can be sampled at a [usually] sub-Nyquist rate of up to $\sim$15 Hz using conventional PIV systems, while data at discrete locations (e.g., unsteady wall pressure) are readily accessible at time-resolved rates. \citet{Tu_2014} showed that it is also possible to resolve the dynamics at a sub-Nyquist sampling rate, which utilized compressed sensing and DMD.  \citet{Luhar2020} recently used rapid distortion theory and Taylor's hypothesis to reconstruct the flowfield in between simulated PIV snapshots from the Johns Hopkins Turbulence Database \citep{Graham:2015}. More commonly, however, in the absence of requisite expensive instrumentation, current experiments employ variants of stochastic estimation \citep{Adrian1979} to estimate the velocity field given statistical information about its relationship to surface pressure or some other time-resolved quantities at a few discrete locations. Modified Stochastic Estimation combines stochastic estimation with POD to estimate the POD temporal coefficients rather than velocity field itself \citep{Taylor:JFE2004, Durgesh2010,Tu2013}. Deficiencies of these stochastic estimation approaches include sensitivity to noise and overfitting (due to a large number of parameters in the  estimation) \citep{Clark:2014jc}. Based on the formulation by \citet{Ewing:1999cd}, \citet{Tinney2006} introduced a novel frequency domain or spectral LSE version that is leveraged in this paper.  
	
	The two approaches herein, called ``SAMMs'', combine spectral LSE from \cite{Tinney2006} with a MIMO system based model and conditioned spectral analysis to obtain POD modes in the frequency (i.e., spectral) domain. SAMMs leverage the ideas proposed by \cite{Lumley1967} concerning space-time POD, and recently \citet{Towne2018} presented an efficient algorithm for application of space-time POD in the frequency domain. \citet{Singh:2020} verified that the algorithm of \citet{Towne2018} yields the same modal structures as previous space-time versions of the POD but in a significantly more computationally efficient manner.
	
	As originally shown by \cite{Lumley1967} and emphasized by \cite{George1988}, POD reduces to Fourier analysis for stationary (i.e., time) or homogeneous (i.e., space) data. Capitalizing on the stationarity assumption, the approach exploits our ability to measure the auto- and cross-correlation functions between spatially- (but not time) resolved flow field data and time- (but not spatially) resolved surface pressure data. Here, we transform these measured functions to the frequency domain using the Fast Fourier Transform (FFT) and employ MIMO conditional spectral analysis methods to determine the transfer functions for the common general case where the input pressure signals are partially correlated. An independent source model combined with an eigendecomposition of the input cross spectral density matrix of the unsteady surface pressures leads to a rapid eigendecomposition of the cross-spectral density matrix of the velocity field.  
	
	In the current study, a canonical cavity flow at Mach 0.6, which exhibits strong self-sustained oscillations, is our targeted flow and is used to demonstrate SAMMs.  The resulting dominant POD modes are compared with those obtained from independent TR-PIV measurements processed using the pwelch algorithm of \citet{Towne2018}.  The primary contribution of this work is a simple method to perform POD in the frequency domain without TR velocity field measurements.
	
	\section{Methodologies}
	\label{sec:method}
	\subsection{Facility and Cavity Model}
	\label{sec:1}
	The experiments are performed in the Pilot Wind Tunnel facility located at the Florida Center for Advanced Aero-Propulsion (FCAAP) at the Florida State University (FSU). This wind tunnel is a blowdown facility with air supplied from 3.4 MPa storage tanks. High-speed subsonic flow (Mach 0.6 for the current experiments) is achieved using a converging nozzle. The stagnation pressure, $p_0$, is measured by a pitot tube located inside the settling chamber using an Omega pressure transducer PX409-050A5V-XL with an uncertainty of $\pm 100$ Pa). The stagnation temperature, $T_0$, is measured by a RTD (uncertainty of $\pm$ 0.1 K) in the settling chamber. The static pressure, $p_s$, is measured using a  pressure transducer (Omega PX303-015A5V, uncertainty of $\pm 250$ Pa) via a pressure tap on the sidewall upstream of the cavity model. Wind tunnel flow conditions are monitored and controlled through a LabVIEW program.
	
\begin{figure}[!htpb]
		\centering
		\begin{subfigure}[b]{0.38\textwidth}
			\includegraphics[width=\textwidth]{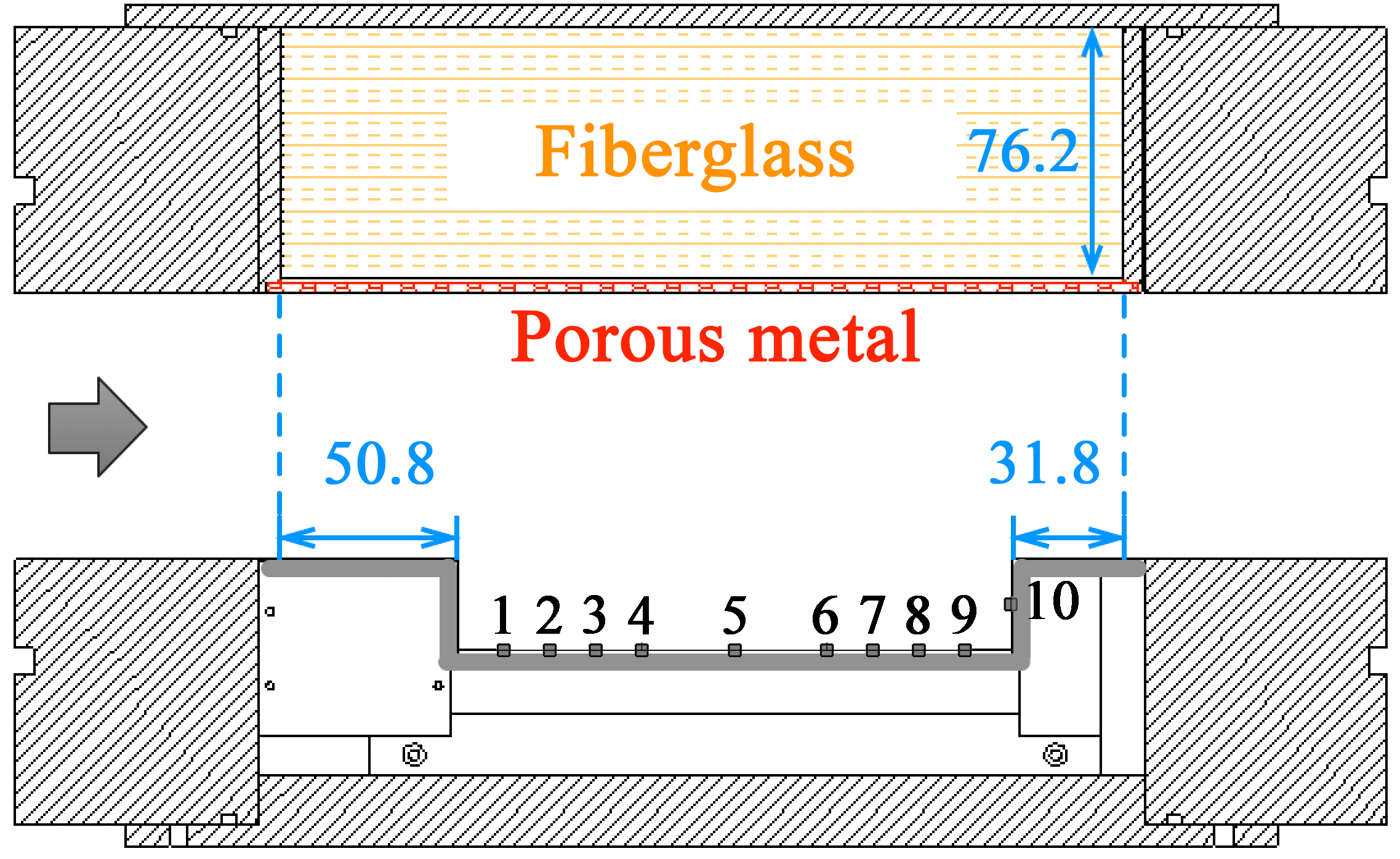}
		\end{subfigure}
		\begin{subfigure}[b]{0.31\textwidth}
			\includegraphics[width=\textwidth]{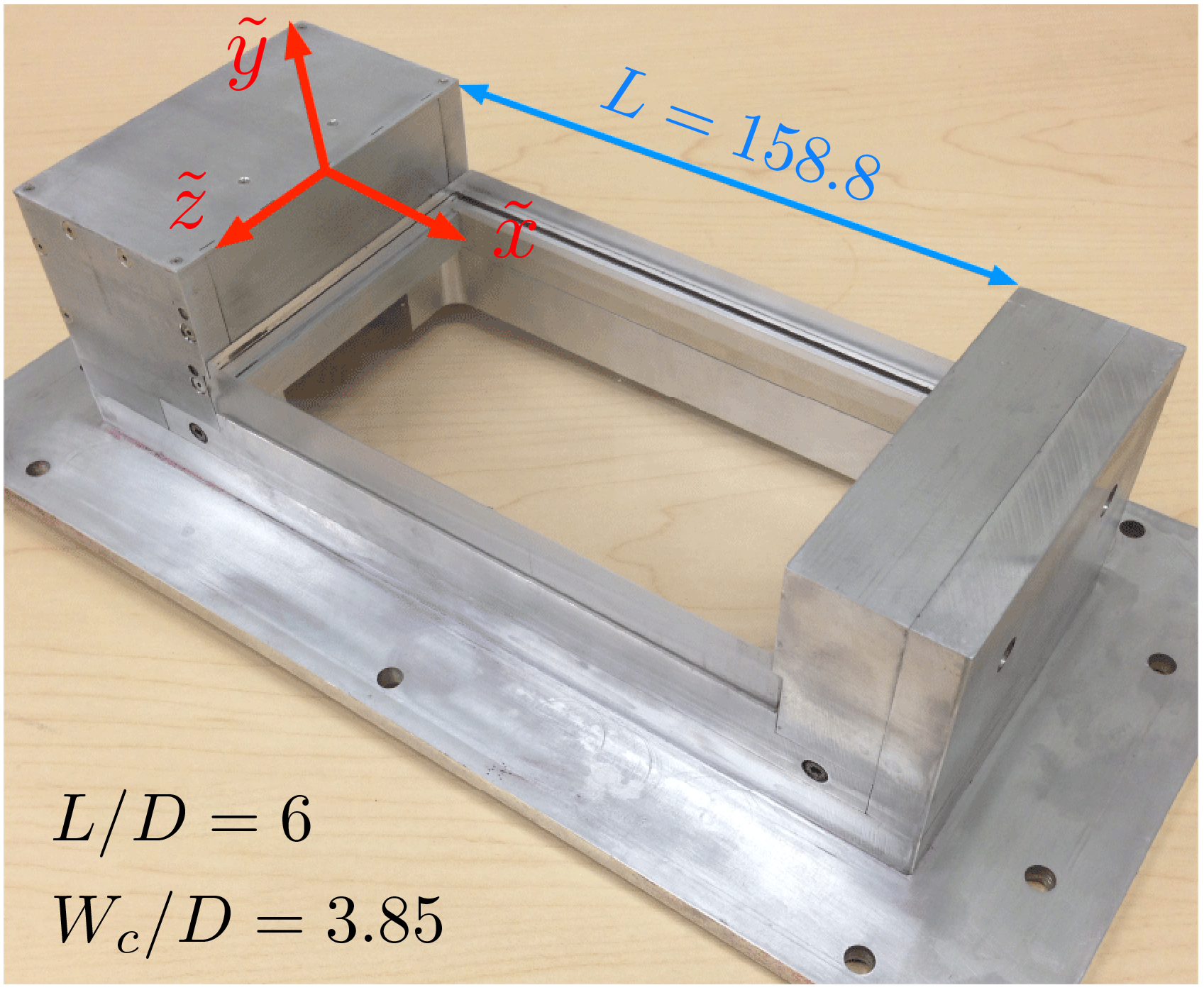}
		\end{subfigure}
		\caption{Schematic of the cavity model (units are in mm). Coordinates system $\tilde x$, $\tilde y$, and $\tilde z$ are non-dimensionalized by $D$.}
		\label{fig:overall}
	\end{figure}

	As shown in Fig.~\ref{fig:overall}, the origin of the coordinate system is fixed at the middle of the cavity leading edge with the $\tilde{x}$, $\tilde{y}$, and $\tilde{z}$ axes denoting the streamwise, normal, and spanwise directions, respectively. The rectangular cavity model has a dimension of $L/D= 6$ and $W_c/D=3.85$ with $D=26.5$ mm. The incoming dimensionless turbulent boundary layer thickness at the cavity leading edge, $\delta_0/D$, is approximately 0.16 with a shape factor of approximately 1.4 at Mach 0.6. The cavity model has an acoustically-treated ceiling (MKI Dynapore P/N 408020 metal sheet and 76.2-mm-thick bulk fiberglass) opposite to the cavity opening. 
	
	\subsection{Synchronized Particle Image Velocimetry and unsteady surface pressure measurements}
	\label{sec:2}
\begin{figure}[!htpb]
		\centering
		\includegraphics[width=0.3\textwidth]{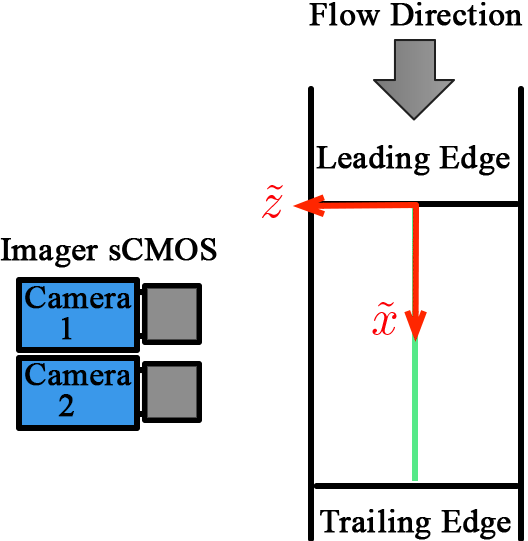}
		\caption{Schematic of 2C2D PIV setup (not to scale).}
		\label{fig:2cpiv}
	\end{figure}

	Two-component, two-dimensional (2C2D) PIV is performed to obtain the streamwise flow field in the $(\tilde{x},\tilde{y})$ plane along the centerline of the cavity. A schematic of the setup is illustrated in Fig.~\ref{fig:2cpiv}. A double-pulse Evergreen Nd:YAG laser (EVG00200) produces laser pulses at a repetition rate of 15 Hz. The beams travel through a plano-convex lens ($f=1000$ mm), a plano-concave cylindrical lens ($f=-9.7$ mm), and then reflect from a folding mirror and through the transparent cavity floor, resulting in a laser sheet with a thickness of approximately $1.5$ mm.  The sheet illuminates nominally 0.3 $\mu$m diameter particles injected upstream of the stagnation chamber using a customized Wright nebulizer seeder (model number 200082000135) with Rosco fog fluid \citep{Alkislar2001}. Two Imager sCMOS cameras with $2560\times2160$ pixels, each equipped with a Nikon Micro-Nikkor 55 mm 1:2.8 lens and a $532$ nm band-pass filter, are oriented with their optical axes normal to the laser sheet. Calibration is performed using a customized 2-D dot pattern. A total number of 4772 snapshots are acquired. Data processing is performed using a multi-pass scheme with a window size ranging from $128\times128$ to $32\times32$ pixels and a 75\% overlap in DaVis 8.3.1, and the resulting vector resolution is 2.8 vectors/mm. 
	
	Nine Kulite sensors are flush mounted in the transparent cavity floor, and a $10^{th}$ sensor is flush mounted in the middle of the aft wall.  Their locations are provided in Table~\ref{tab:10kulite} and illustrated in Fig. \ref{fig:overall}. All the sensors on the floor are slightly shifted from the mid-plane to transmit the laser sheet. Unsteady surface pressure measurements are acquired at a sampling rate of 204.8 kHz for 15 seconds for each PIV run using a NI PXI-1031 chassis and three NI PXI-4462 24 bits cards. A Scientech photodiode signal is acquired simultaneously with the unsteady pressure signals during the PIV measurements to indicate the laser pulse instances in the data reduction. The unsteady pressure data are subsequently low-pass filtered and downsampled by a factor of 8 to an effective sampling rate of 25.6 kHz.
	
	\begin{table}[!htpb]
		\centering
		\caption{Kulite locations for velocity-pressure coupling measurements}
		\label{tab:10kulite}
		\small
		\begin{tabular}{c c c c c}
			\hline\hline
			Kulite model & Index & $\tilde{x}$ & $\tilde{y}$ & $\tilde{z}$ \\\hline
			XCQ062-5D    & 1              & 0.5   & -1    & -0.19  \\
			XCS062-10D   & 2              & 1     & -1    & -0.19  \\
			XCQ062-10D   & 3              & 1.5   & -1    & -0.19  \\
			XCQ062-10D   & 4              & 2     & -1    & -0.19  \\
			XCS062-10D   & 5              & 3     & -1    & -0.19  \\
			XCS062-10D   & 6              & 4     & -1    & -0.19  \\
			XCS062-10D   & 7              & 4.5   & -1    & -0.19  \\
			XCQ062-1BAR  & 8              & 5     & -1    & -0.19  \\
			XCQ062-1BAR  & 9              & 5.5   & -1    & -0.19  \\
			XT190-25PSIA & 10             & 6     & -0.5  & 0   \\\hline\hline
		\end{tabular}
	\end{table}
	
	\subsection{TR-PIV measurements}
	Independent TR-PIV measurements are obtained to provide a comparison at the same flow condition. The TR-PIV measurements use a Photonics DM dual-head laser and a Phantom V2012 high speed camera. A 105 mm f/2.8 Nikon lens with a 532 nm band pass filter is used in this case. The data were taken at a sampling rate of 16 kHz for one second in double-frame mode with an image resolution of $1280 \times 464$ pixels. At this sampling rate, the velocity field associated with the first four Rossiter frequencies below approximately 2 kHz are temporally resolved. The field of view covers the leading edge to trailing edge of the cavity to maximize pixel resolution and avoid the peak locking effects. A 96$\times$96 to 32$\times$32 multi-pass scheme with a $75\%$ overlap is used to calculate the velocity field, resulting in a vector resolution of approximately 1 vector/mm. It should be noted that both NTR and TR-PIV data were post-processed by universal-outlier-detection \citep{Westerweel2005} in DaVis and Multivariate outlier detection \citep{Griffin2010} using MATLAB. 
	
	\subsection{Multi-input/multi-output system}
	\label{sec:MIMO}
\begin{figure}[tbph!]
		\centering
		\includegraphics[width=0.3\textwidth]{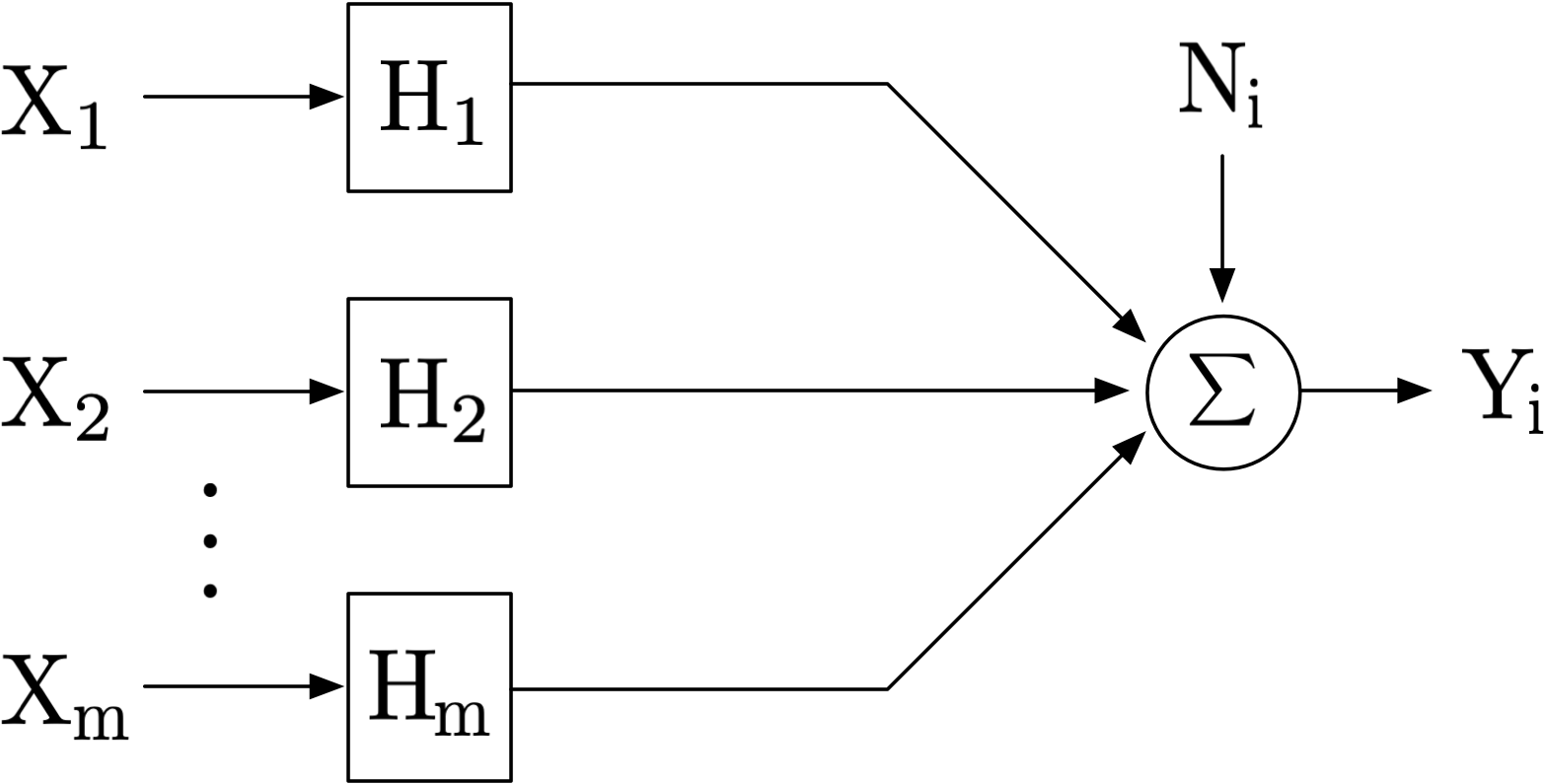}
		\caption{Schematic of a MISO system. A similar model is constructed at each output measurement location to produce a MIMO model.}
		\label{fig:mimo}
	\end{figure}

	A MIMO model is devised to describe the relationship between the $m$ unsteady surface pressure inputs and the output $(u,v)$ velocity at $n$ measurement locations \citep{BP}.  Fig. \ref{fig:mimo} illustrates a multi-input/single-output (MISO) model.  A MIMO model is simply a MISO model for each velocity component at each measurement location.
	
	It should be noted that the model is linear and approximates the nonlinear flow system. The velocity at each measurement location of interest is a sum of the contributions from all sensors and noise that is unique to each output and uncorrelated with the inputs.
	
	By performing a Fourier transform, the time-domain model is transformed to the frequency domain (denoted by capital letters)
	\begin{equation}\label{eq:model}
	Y=HX + N,
	\end{equation}
	where $X \in \mathbb{C}(m \times 1)$, $H \in \mathbb{C}(2n \times m)$, and both $Y$ and $N \in \mathbb{C}(2n \times 1)$.  Explicit frequency dependence in Eq. \ref{eq:model} and below has been suppressed.  Note that the inputs are, in general, partially correlated.  We wish to determine the optimal $H$ that minimizes the output noise contribution. Solving for $H$ generally requires TR pressure and velocity data. Unfortunately, TR velocity is not available in a conventional PIV system due to the low sampling frequency.
	
	Right multiplying Eq. \ref{eq:model} by the complex conjugate transpose $X'$ and taking $\frac{2}{T} E[^.]$, where $E[^.]$ denotes the expectation operator, yields
	\begin{equation}
	G_{xy}=H G_{x_ix_j}, \label{eq:H}
	\end{equation}
	where $G_{xy} \in \mathbb{C}(2n \times m)$ is the cross spectral density matrix between pressure and velocity, and $G_{x_ix_j} \in \mathbb{C}(m \times m)$ is the Hermitian cross spectral density matrix of pressure, and whose diagonal elements are the autospectral density of the inputs. In general, $G_{x_ix_j}$ is full rank, enabling matrix inversion of Eq. \ref{eq:H} to solve for $H$ at each frequency of interest. However, the equation above still cannot be calculated directly without TR velocity. Therefore, an alternative approach uses
	\begin{equation}
	G_{xy}(f)=\int_{-\infty}^{+\infty}R_{xy}(\tau)e^{-j2\pi f\tau}d\tau,
	\end{equation}
	where $R_{xy}$ is the cross-correlation matrix between pressure fluctuations, $x$, and velocity fluctuations, $y$, which can be calculated from the TR pressure data and the non-time-resolved (NTR) velocity by systematically shifting the time delay between pressure and velocity snapshots. It should be noted that this methodology is equivalent to that employed in the Spectral LSE proposed by \cite{Tinney2006}, but our implementation is explained below for clarity and completeness.
	
	For stationary data, the cross-correlation $R_{xy}$ is defined as
	\begin{equation}
	R_{xy}(\tau)=E[x(t)y(t+\tau)].
	\end{equation}
	However, 
	\begin{equation}
	R_{xy}(\tau)=R_{yx}(-\tau)=E[y(t)x(t-\tau)].
	\label{eq:Rxy}
	\end{equation}
	The last term in Eq. \ref{eq:Rxy} is used to compute the cross correlation between velocity $y$ and pressure $x$ by sequentially shifting the pressure signal with respect to the PIV snapshots and averaging over all snapshots.  Then the single-sided cross-spectral density is calculated via a Discrete Fourier Transform of $R_{xy}(\tau)$
	
	\begin{equation} \label{eq:Gxy}
	\begin{split}
	{G_{xy}}\left( {{f_j}} \right) & = c\Delta t\sum\limits_{k = 1}^{N_{FFT}} {w(k){R_{xy}}\left( k \right){e^{ - i\frac{{2\pi \left( {j - 1} \right)\left( {k - 1 - \frac{N_{FFT}}{2}} \right)}}{N_{FFT}}}}}; \\
	j & = 1:N_{FFT}/2 + 1,
	\end{split}
	\end{equation}
	where ${f_j}=(j-1)f_s/N_{FFT}$, $w$ is a window function, $c=1$ for $j=1$ and $c=2$ for $j \geq 2$, and ${R_{xy}}\left( k \right) = {R_{xy}}\left( {{{ - N_{FFT}\Delta t} \mathord{\left/
				{\vphantom {{ - N_{FFT}\Delta t} {2 + \left( {k - 1} \right)\Delta t}}} \right.
				\kern-\nulldelimiterspace} {2 + \left( {k - 1} \right)\Delta t}}} \right)$. By factoring and simplifying the phase term,
	\begin{equation}
	{e^{ - i\frac{{2\pi \left( {j - 1} \right)\left( { - {N_{FFT} \mathord{\left/
								{\vphantom {N_{FFT} 2}} \right.
								\kern-\nulldelimiterspace} 2}} \right)}}{N_{FFT}}}} = {e^{i\pi \left( {j - 1} \right)}},
	\end{equation}
	we leverage the Fast Fourier Transform (when $N_{FFT}$ is a power of 2), so that
	\begin{equation}
	G_{xy}\left( {{f_j}} \right) = c\Delta t{e^{i\pi \left( {j - 1} \right)}}\,FFT(w \cdot {R_{xy}},N_{FFT}).
	\label{eq:GxyFFT}
	\end{equation}
	
	After obtaining $G_{xy}$ and $G_{xx}$, $H$ is determined using Eq. \ref{eq:H}.
	
	Note that the POD eigenvectors $\Phi$ and eigenvalues $\Lambda$ can be obtained via solution of the $2n \times 2n$ eigenvalue problem
	\begin{equation}
	\frac{YY'}{K}\frac{I_{2n}}{2n}\Phi  = \Phi \Lambda,
	\label{eq:Gyy_norm}
	\end{equation}
	where $I_{2n}/(2n)$ is a normalized area weight factor $=dA/A$ used to enable the comparison between different 2C2D PIV grids of aggregate size $n$, and $I_{2n}$ is the $2n \times 2n$ identity matrix. This eigenvalue problem is large and intractable for PIV data on current desktop computers. As described in \citet{SchmidtSPOD}, the eigenvectors $\Phi$ are linear combinations of the velocity snapshots $Y$, leading to the (typically) smaller $K \times K$ eigenvalue problem via the method of snapshots, where $K$ is the number of blocks \citep{Sirovich1987}
	\begin{equation}
	\frac{Y' Y}{2n K}\Psi  = \Psi \Lambda, \quad \Phi  =  Y\Psi,
	\label{eq:method_of_snapshot}
	\end{equation}
	where the modes $\Phi$ are normalized by their 2-norm such that $\Phi'\Phi/(2n)=I$. This approach is denoted as ``SAMM-SPOD'' because it uses the SPOD algorithm of \citet{Towne2018} as follows.  After $H$ is determined from Eq. \ref{eq:H}, Eq. \ref{eq:model} is used to generate the estimated data matrix $\hat{Y}=HX$ (ignoring the noise term) from the FFT of the windowed and, if applicable, overlapped time-resolved pressure data records $X$. The details of this implementation are described in \citet{Zhang2019}.  It should be noted that acquiring the pressure data as long continuous records offers the possibility of overlapping in the construction of the data matrix. In the follow section, we introduce a simpler way to estimate the dominant POD modes and eigenvalues.
	
	\subsection{Conditional Spectral Analysis}
\begin{figure}
		\centering
		\includegraphics[trim = 75mm 72mm 85mm 45mm, clip,width=3in]{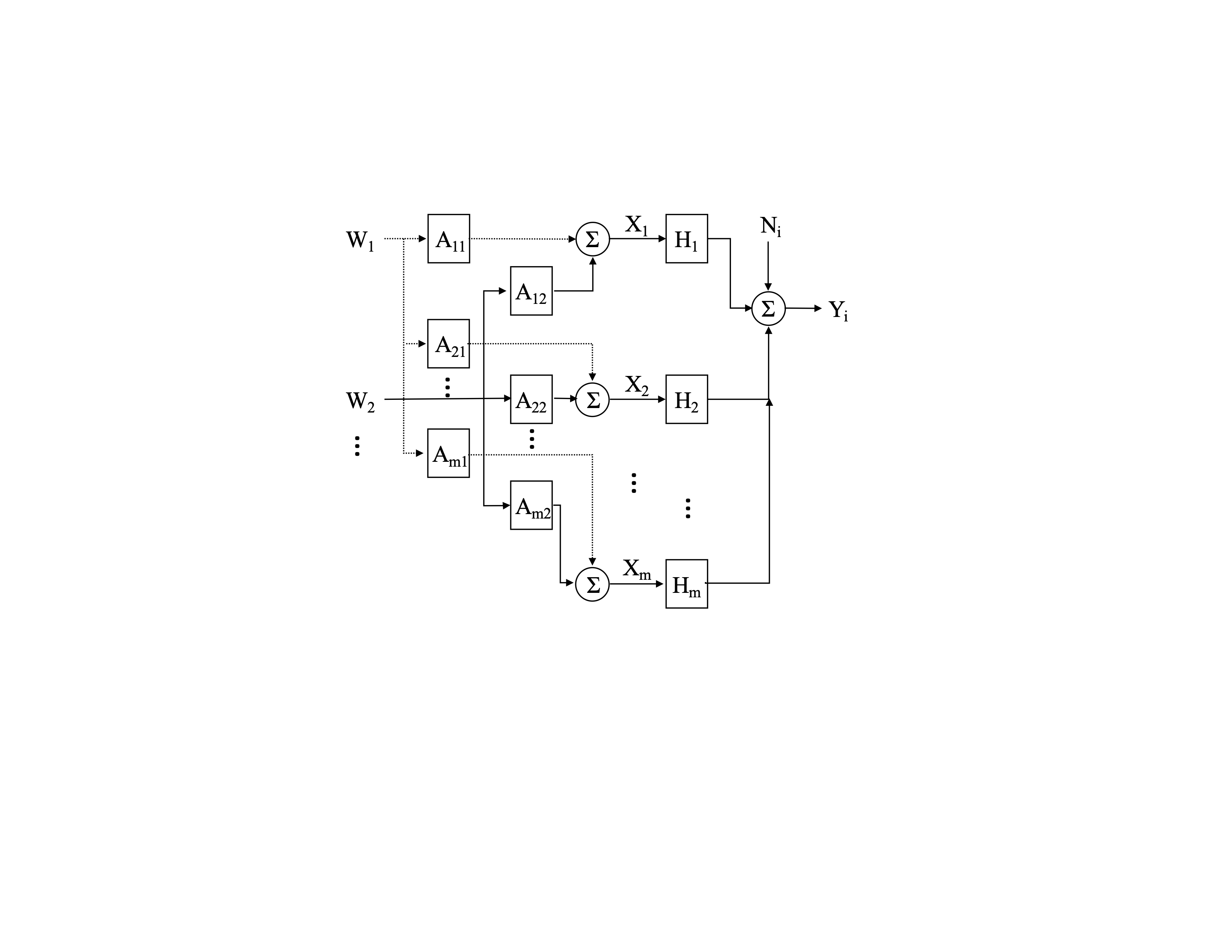}
		\caption{Schematic of a MIMO system in the frequency domain shown with two independent inputs.}
		\label{fig:indinput}
	\end{figure}

	The conditional spectral analysis herein is adapted from \cite{BP}. The objective of the analysis is to determine the relationship between \emph{unknown} independent inputs $W$ and the outputs, $Y$, as depicted in Fig. \ref{fig:indinput}. Right multiplying Eq. \ref{eq:model} by the complex conjugate transpose $Y'$ and taking $\frac{2}{T} E[^.]$ of both sides yields the cross spectral density matrix of the velocity
	\begin{equation}
	G_{y_iy_j} = HG_{yx}+G_{nn},
	\label{eq:Gyy_nn}
	\end{equation}
	where the first term on the right-hand side represents the model output, $\hat{G}_{{y_i}{y_j}}$, and the second term is an unknown diagonal matrix representing the noise. From Eq. \ref{eq:H} and noting that $G_{xy}=G_{yx}'$, the model output can be expressed as
	\begin{equation}
	\hat{G}_{{y_i}{y_j}} = HG_{x_ix_j}'H'.
	\label{eq:Gyy}
	\end{equation}
	Eq. \ref{eq:Gyy} provides a way to compute the estimated cross-spectral density of the velocity field.
	
\begin{figure}[!htpb]
		\centering
		\includegraphics[width=0.48\textwidth]{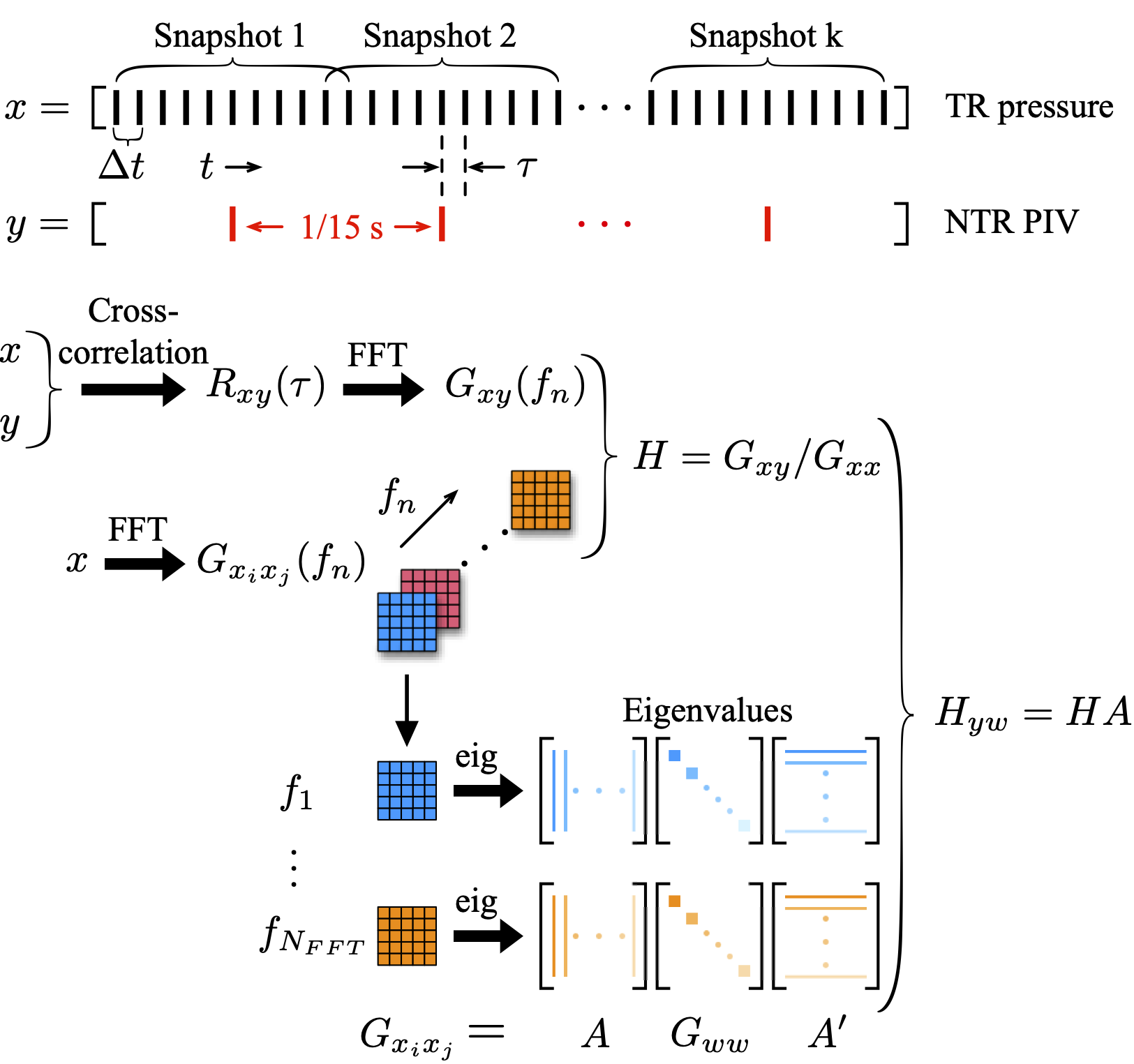}
		\caption{The ``SAMM-RR'' algorithm.}
		\label{fig:algorithm}
	\end{figure}

	As noted earlier, the pressure inputs $X$ are partially coherent and therefore not independent. Now assuming $X$ is a linear combination of $r \leq m$ independent inputs $W$ via a transfer matrix $A\in \mathbb{C}(m\times m)$ as shown in Fig. \ref{fig:indinput}, then
	\begin{equation}
	X=AW.
	\label{eq:w}
	\end{equation}
	Right multiplying this equation by $X'$ and taking $\frac{2}{T} E[^.]$ yields
	\begin{equation}
	\frac{2}{T}E[XX'] = \frac{2}{T}E[AW(AW)'] = A\frac{2}{T}E[WW']A',
	\end{equation}
	or
	\begin{equation}
	G_{x_ix_j} = AG_{ww}A'.
	\label{eq:Gww}
	\end{equation}
	Therefore, a sorted eigendecomposition of the \emph{small} $m \times m$ Hermitian input cross-spectral density matrix yields $A$ and the autospectral density of the independent sources $W$ in descending order of the eigenvalues and corresponding eigenmodes.  Substituting Eq. \ref{eq:Gww} into Eq. \ref{eq:Gyy} yields
	\begin{align} \label{eq:HGH}
	\begin{split}
	\underbrace{\hat{G}_{y_iy_j}}_{2n\times 2n} &= \underbrace{H}_{2n\times m}\underbrace{A}_{m\times m}\underbrace{G_{ww}}_{m\times m}\underbrace{A'}_{m\times m}\underbrace{H'}_{m\times 2n} 
	\end{split}
	\end{align}
	or 
	\begin{equation} \label{eq:Gyy_reduced}
	\hat{G}_{y_i y_j} = H_{yw} G_{ww} H_{yw}',
	\end{equation}
	which is the desired reduced-rank approximation of the cross spectral density matrix of the velocity field, denoted as ``SAMM-RR''.  Note that $H_{yw}= HA$ is the estimated frequency response function between the model output velocity $\hat{Y}$ and independent inputs $W$ and effectively provides the response modes of a resolvent analysis \citep{Taira_2017}. Using the same scaling as in Eq. \eqref{eq:Gyy_norm} yields
	\begin{align} \label{eq:Hyw_scale}
	\hat{G}_{y_i y_j} = \overline{H}_{yw}\overline{G}_{ww}\overline{H'}_{yw},
	\end{align}
	where  \textcolor{black}{$\overline{H}_{yw}=\frac{\sqrt{2n}H_{yw}}{\left |H_{yw}\right | }$} and  \textcolor{black}{$\overline{G}_{ww}= \frac{G_{ww}}{2n}\parallel H_{yw} \parallel^2$}. Column $j$ of $\overline{H}_{yw}$ is the estimated POD mode $j$. SAMM-RR is a computationally ``simple'' way to extract the spatiotemporal coherent structures of the velocity fields associated with particular frequencies.  We emphasize that the approaches presented are achieved without TR-PIV. Therefore, the noise term, $G_{nn}$, of Eq. \ref{eq:Gyy_nn} cannot be computed, and so the multiple coherence function normally used to assess model adequacy cannot be evaluated. 
	The SAMM-RR algorithm for calculating the velocity modes is illustrated in Fig. \ref{fig:algorithm} and described as follows
	\begin{enumerate}
		\item Non-dimensionalize the NTR velocity fluctuations and TR surface pressure fluctuations by freestream velocity fluctuation vector $y=\frac{\Vec{u'}}{U_{\infty}}$ and dynamic pressure $x=\frac{p'}{0.5\rho {U_{\infty}}^2}$.
		\item Calculate the input cross spectral density matrix $G_{x_i x_j}$. 
		\item Perform a sorted eigendecomposition of $G_{x_i x_j}$ to obtain the eigenvectors $A$ and eigenvalues $G_{ww}$ at each frequency. 
		\item Calculate the cross-correlation of $x$ and $y$ as $R_{x_i y}$ for each velocity component at each PIV grid point using Eq. \ref{eq:Rxy}. 
		\item Calculate the cross-spectra matrix $G_{x_i y}$ using Eq. \ref{eq:Gxy} with the same frequency resolution and window function as in step 2. 
		\item Calculate $H$ using Eq. \ref{eq:H} for the specified frequencies of interest. 
		\item Calculate the transfer function $H_{yw}$ using Eq. \ref{eq:Gyy_reduced} for the specified frequencies of interest, and then use Eq. \ref{eq:Hyw_scale} for re-scaling. The columns of $\overline{H}_{yw}$ are the low-rank approximation of the spectral POD velocity modes, and $\overline{G}_{ww}$ is the independent source autospectral density matrix \label{step:7}.
	\end{enumerate}
	The application of the above methods to flow-induced cavity oscillations is discussed in the next section.  We compare the non-TR SAMMs presented above to the results from independent TR-PIV measurements.
	\section{Results and Discussion}
	\subsection{Conditioned spectral analysis}
\begin{figure}[!htpb]
		\centering
		\includegraphics[width=0.45\textwidth]{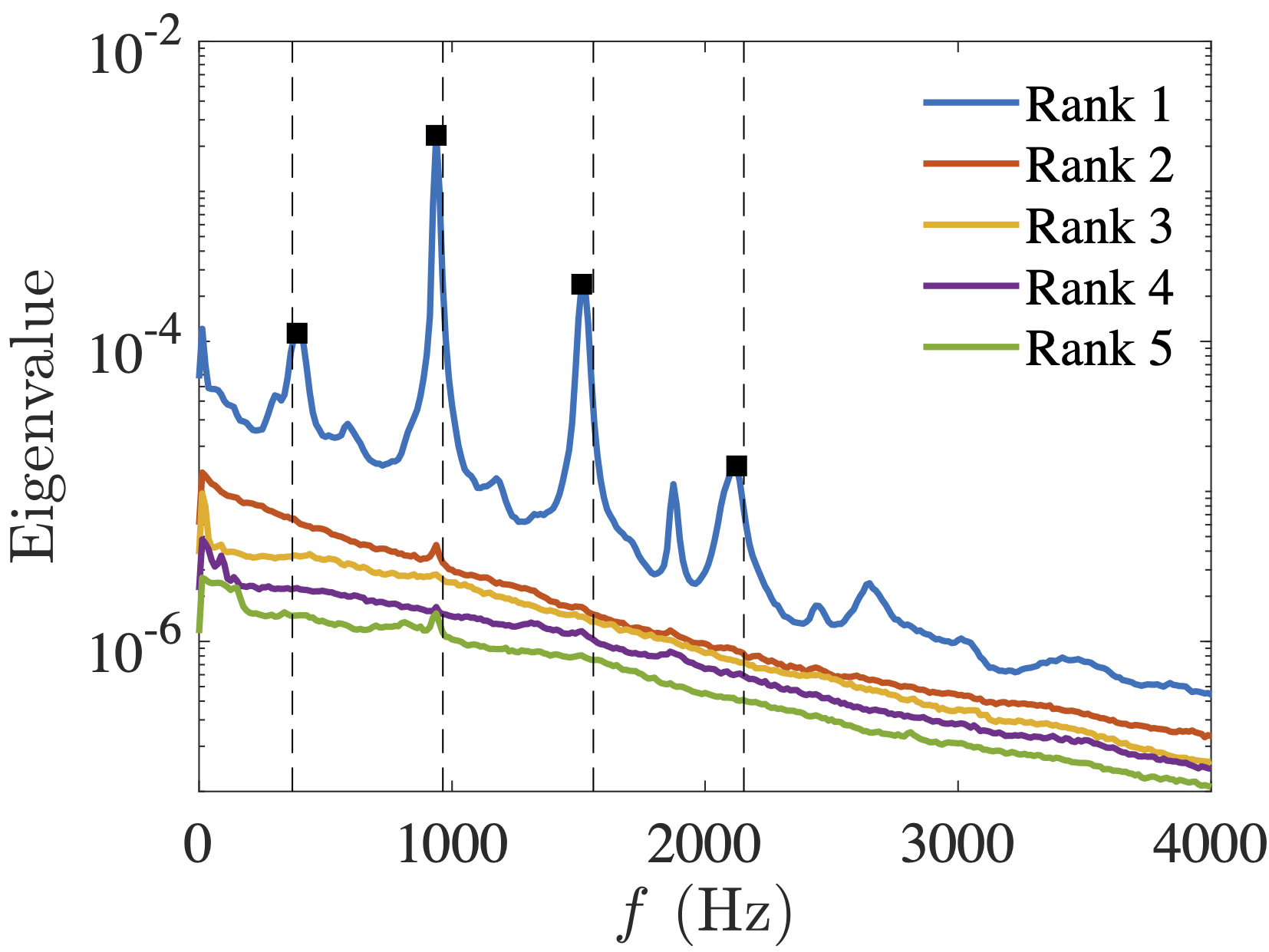}
		\caption{Eigenvalue spectra of the first five modes of $G_{x_ix_j}$ versus frequency using all 10 sensors. The dashed-lines are the predicted Rossiter frequencies using Eq. \ref{eq:Rossiter}. The square markers indicate the frequencies for the velocity modes.}
		\label{fig:SAMM_eigenvalue}
		\vspace{-0.2in}
	\end{figure}

	$G_{x_ix_j}$ consists of the auto- and cross-spectral densities of the pressure signals along and off the diagonal, respectively. As a result of finite random measurement noise, the inputs are partially correlated, in which case $G_{x_ix_j}$ is a full-rank matrix and can be inverted to solve for $H$ via Eq. \ref{eq:H}.  Furthermore, it can be decomposed via an eigendecomposition. Thus, Eq. \ref{eq:Gww} is used to compute the independent source auto-spectral densities, $G_{ww}$, and the eigenvalue spectra of the first five modes are shown in Fig. \ref{fig:SAMM_eigenvalue}.  By construction, the eigenvalues of POD mode 1 possess the well-known characteristics of a cavity flow, showing multiple spectral peaks indicative of the Rossiter tones. The modified Rossiter equation \citep{Heller1971} is
	
	\begin{equation}\label{eq:Rossiter}
	St = \frac{fL}{U_\infty}=\frac{n-\alpha}{1/\kappa+Ma/\sqrt{1+(\gamma-1)Ma^2/2}},
	\end{equation}
	where $\alpha=0.38$ is the phase lag, $\kappa=0.65$ is the convective speed ratio as reported in \cite{ZhangAIAA2019}, $n$ is the Rossiter mode index, and $\gamma$ is the specific heat ratio. It is clear that the prediction matches well with the tonal frequencies. The eigenvalues of lower ranked modes are significantly lower, adding broadband content. The POD velocity modes associated with the Rossiter frequencies, indicated by the black squares, are presented later.
	
	\begin{figure}[!htpb]
		\centering
		\includegraphics[width=0.45\textwidth]{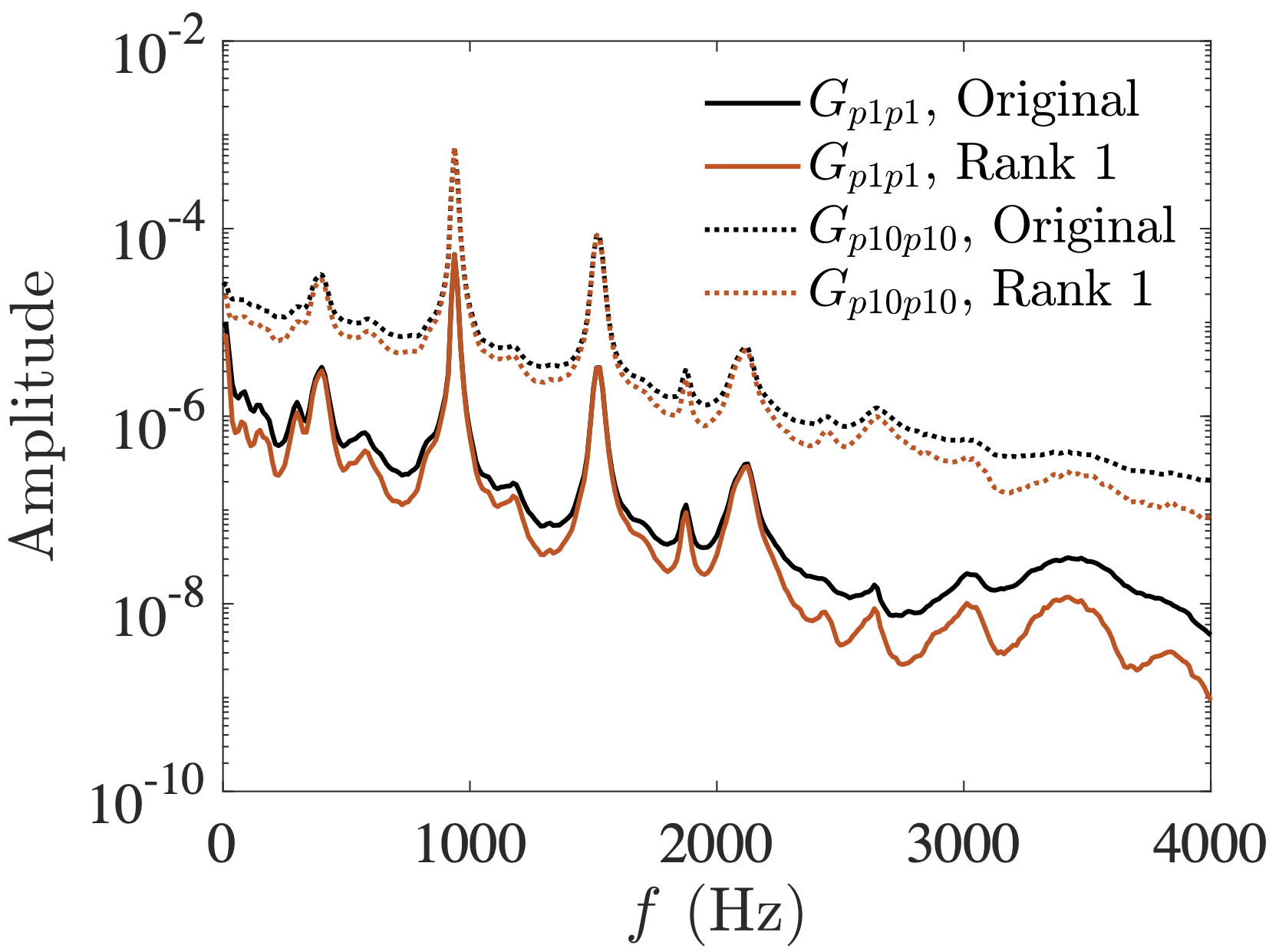}
		\caption{Low-rank approximation of $G_{p_{i}p_{i}}$ using 10 sensors.}
		\label{fig:low_rank_Gxx}
	\end{figure}
	
	Recall that the eigenvectors, $A$, of $G_{x_ix_j}$ are also obtained via Eq. \eqref{eq:Gww}. Therefore, we can construct a low-rank approximation of $G_{x_ix_j}$. Rank-1 approximations of $G_{x_ix_i}$ of sensors 1 and 10 are compared to their respective measured autospectral density in Fig. \ref{fig:low_rank_Gxx}. The rank-1 approximations match well with the measured autospectral density at the narrowband peaks. Minor discrepancies exist in the broadband content, which shows that broadband turbulent features cannot be accurately captured with such an approximation. Fortunately, we are primarily interested in capturing the dynamics of the energetic coherent flow structures, so the rank-1 approximation is suitable in this instance.
	
	\subsection{Spectral POD modes}
	With the TR-PIV data, the spectral POD modes of the velocity fields can be directly calculated using the pwelch algorithm provided by \cite{Towne2018} with the same normalized area weighting approach as for the 15 Hz PIV data. A Hamming window is used, and the $N_{FFT}$ is 1280 with a 75\% overlap, which results in 23 effective blocks. The random uncertainty associated with this analysis is approximately 20\% due to the limited number of records. With these settings, the spectral resolution is 12.5 Hz, which is chosen to match that of the SAMMs with $N_{FFT}=2048$.
	
	\begin{figure}[!htpb]
		\centering
		\includegraphics[width=0.45\textwidth]{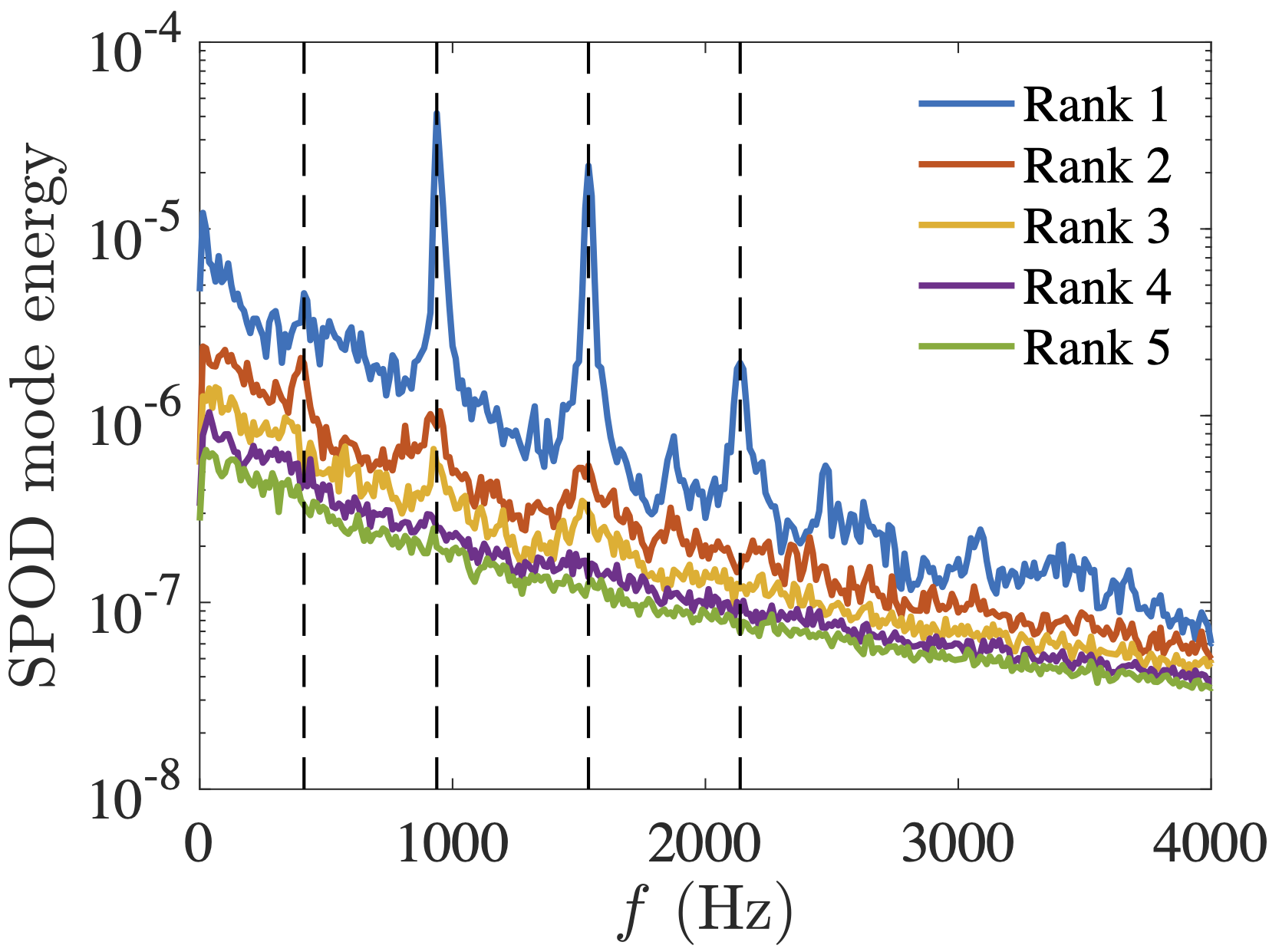}
		\caption{Eigenvalue spectra of the first five modes versus frequency from TR-PIV SPOD.}
		\label{fig:SPOD_energy}
	\end{figure}
	
\begin{figure*}[!htpb]
		\centering
		\begin{subfigure}[b]{1\textwidth}
			\centering
			\includegraphics[width=\textwidth]{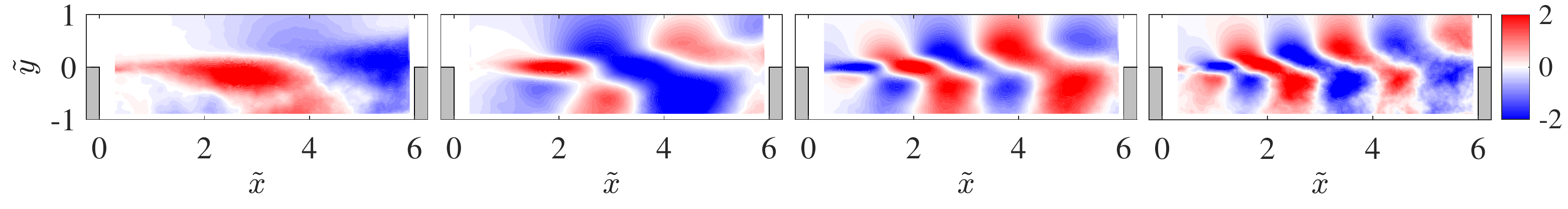}
			\vspace{-10pt}
			\caption{Mode shape of $u$}
			\vspace{0pt}
		\end{subfigure}
		\begin{subfigure}[b]{1\textwidth}
			\centering
			\includegraphics[width=\textwidth]{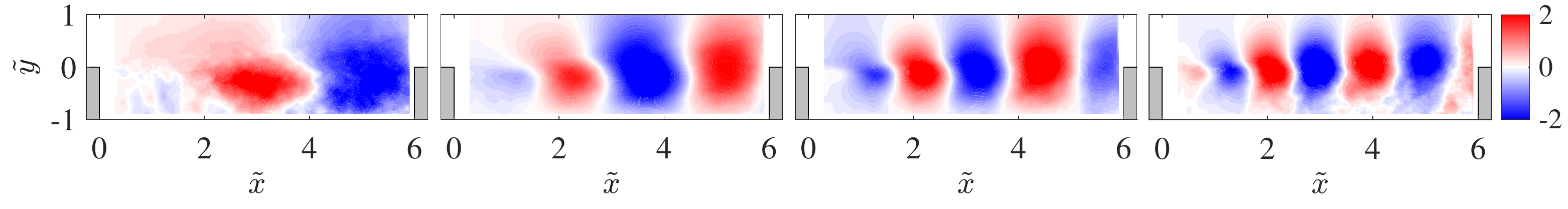}
			\vspace{-10pt}
			\caption{Mode shape of $v$}
		\end{subfigure}
		\caption{Dominant mode shapes corresponding to first 4 Rossiter frequencies from SPOD of TR-PIV.}
		\label{fig:SPOD_modes}
	\end{figure*}

	Fig. \ref{fig:SPOD_energy} shows the re-scaled eigenvalue spectra for the five highest-ranked POD modes. The energy in the rank-1 mode is significantly higher than the rest, which is similar to that observed for $G_{x_ix_j}$ in the previous section. The modes shapes associated with the Rossiter frequencies indicated by the dashed lines are extracted as shown in Fig. \ref{fig:SPOD_modes}. Mode shapes of both $u$ and $v$ exhibit traveling wave patterns that are clearly observed via animation of the modes (not shown). The wavelength of the coherent structures reduces with increasing frequency as expected.
	
	\subsubsection{SAMM-SPOD}
	For SAMMs, one of the open questions is the number and location of unsteady pressure sensors used in the MIMO system \citep{Srini:2007}. In an open cavity, in which the shear layer spans the cavity mouth, the trailing edge is generally acknowledged as the acoustic source that generates acoustic waves, indicative of a single dominant source. Fig. \ref{fig:SAMM_eigenvalue} verifies that only one independent source is required to describe the characteristics of the cavity flow. It is perhaps intuitive to choose at least one sensor near the trailing edge. Therefore, we tested three combinations of sensors to investigate the effects: 
	\begin{enumerate}
		\item All ten sensors,
		\item Two sensors with one near the leading edge (sensor 1) and the sensor on the aft wall (sensor 10), and 
		\item Two sensors near the leading edge (sensors 1 and 2, see Fig. \ref{fig:overall}).
	\end{enumerate} 
	
	Case 1 uses information from all sensors, while Case 2 captures the global features with just two sensors closest to the source. Case 3 is challenging because these two sensors are located farthest from the source. The following discussion compares these three cases.
	
\begin{figure}[!htpb]
		\centering
		\includegraphics[width=0.45\textwidth]{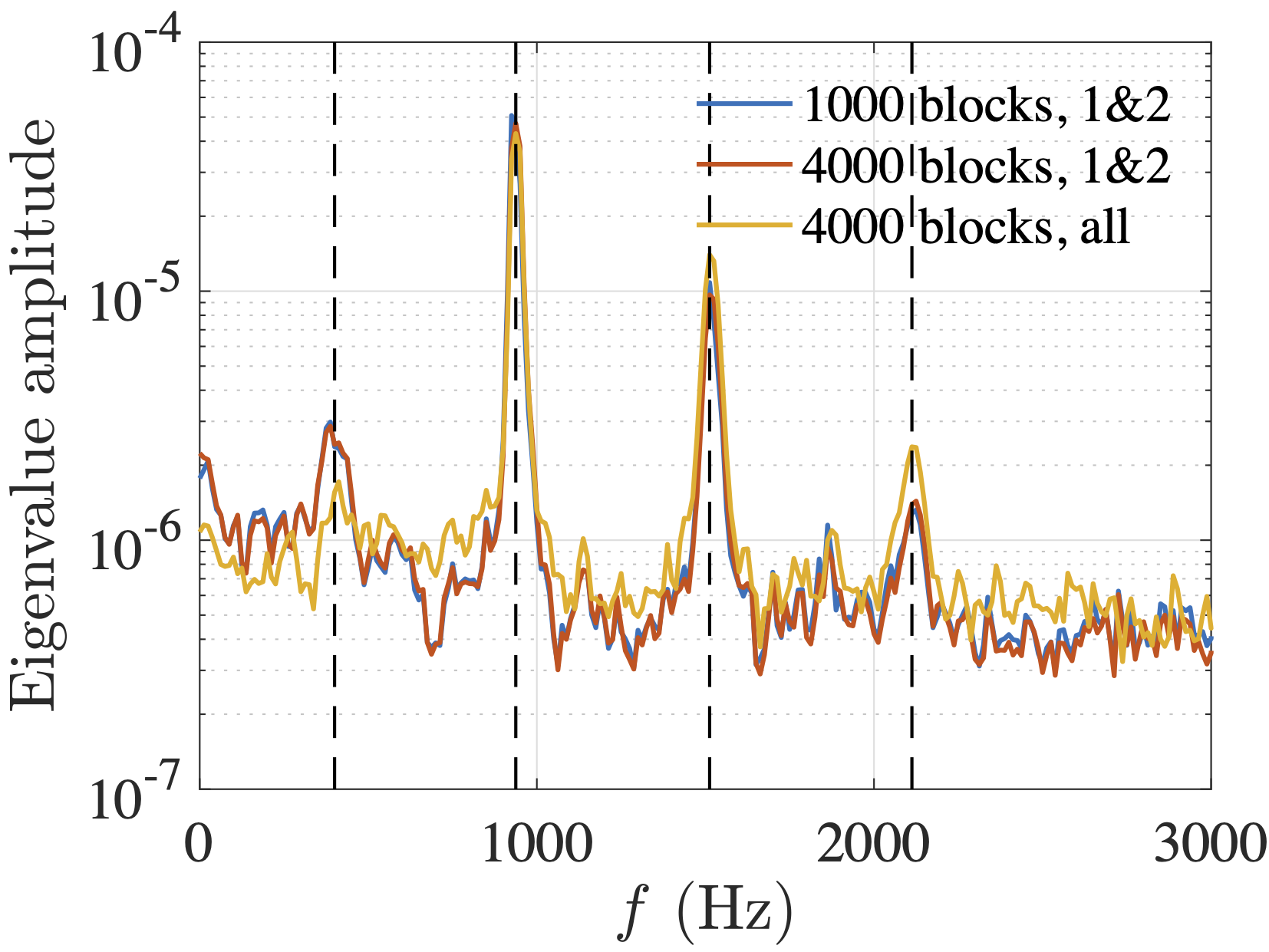}
		\caption{Eigenvalues from SAMM-SPOD for $v$-component versus frequency. The dashed lines indicate Rossiter frequencies.}
		\label{fig:SAMM_SPOD_eigenvalues}
	\end{figure}

	As described previously, SAMM-SPOD calculates $X$ for multiple blocks and $\hat{Y}'\hat{Y}$ directly after computing $H$.  The POD modes are then obtained by solving the smaller snapshot eigenvalue problem using Eq. \ref{eq:method_of_snapshot}. The spectra of the rank-1 eigenvalues for different cases are compared in Fig. \ref{fig:SAMM_SPOD_eigenvalues}. For the same combination of sensors (1 and 2), the eigenvalues do not change significantly with an increasing of number of blocks. This is expected as the eigenvalues are the characteristics of the system. When all of the sensors are used, the eigenvalue spectrum changes slightly at the less-energetic first and fourth Rossiter frequencies. The eigenvalue amplitude does not change significantly at the most energetic ($2^{nd}$ and $3^{rd}$) Rossiter frequencies. 
	
\begin{figure*}[!htpb]
		\centering
		\begin{subfigure}[b]{1\textwidth}
			\centering
			\includegraphics[width=\textwidth]{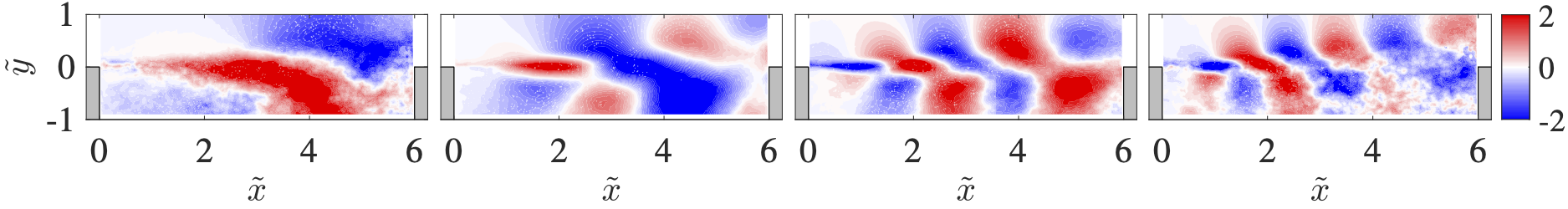}
			\vspace{-12pt}
			\caption{mode shape of $u$}
			\vspace{0pt}
		\end{subfigure}
		\begin{subfigure}[b]{1\textwidth}
			\centering
			\includegraphics[width=\textwidth]{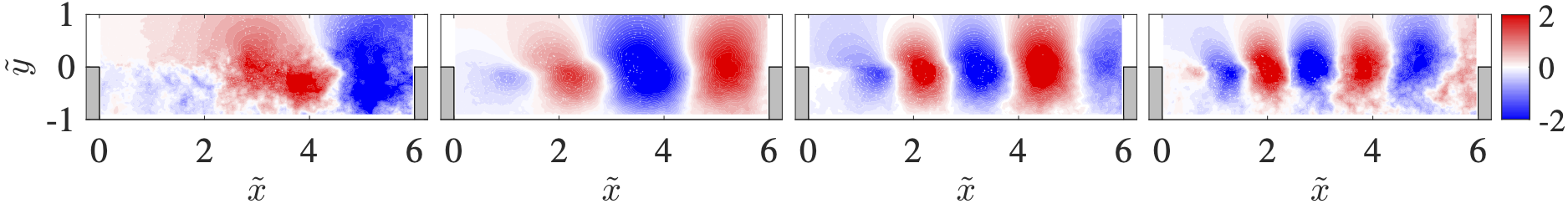}
			\vspace{-12pt}
			\caption{mode shape of $v$}
			\vspace{0pt}
		\end{subfigure}
		\caption{Rank-1 POD modes corresponding to first 4 Rossiter frequencies from SAMM-SPOD using 10 sensors.}
		\label{fig:SAMM_SPOD_modes}
	\end{figure*}

	With all 4772 blocks used in the SAMM-SPOD in the current example to obtain the modes, the results are shown in Fig. \ref{fig:SAMM_SPOD_modes}. We find that the mode shapes are consistent for the different sensor combination cases; therefore, only the mode shapes for the all-sensors case are shown for brevity. We note that increasing the number of blocks reduces the noise in the ensemble average. The modes show, suitably adjusted in phase, are in very good agreement with those from the TR-PIV results shown in Fig. \ref{fig:SPOD_modes}, particularly the dominant second and third Rossiter modes.
	
	\subsubsection{SAMM-RR}
\begin{figure}[!htpb]
		\centering
		\includegraphics[width=0.45\textwidth]{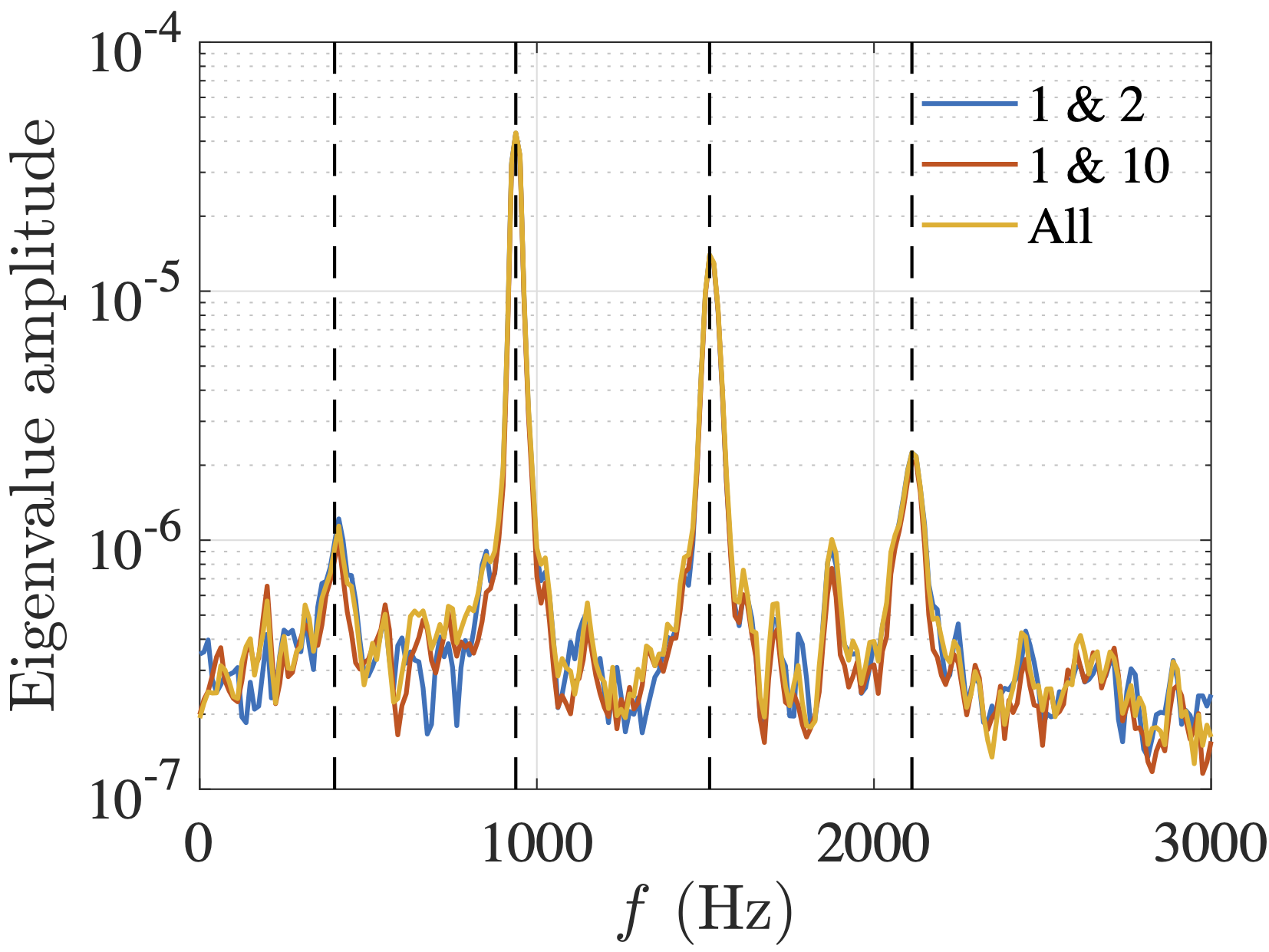}
		\caption{Eigenvalues (rank-1 of $\overline{G}_{ww}$) for $v$-component versus frequency. The dashed lines indicate the Rossiter frequencies.}
		\label{fig:SAMM_eigenvalue_2sensors}
	\end{figure}

	The SAMM-RR algorithm shown in Fig. \ref{fig:algorithm} is a reduced-rank approximation that is much more computationally efficient than SAMM-SPOD since it does not require an eigendecomposition of the velocity cross-spectral matrix. The re-scaled eigenvalues ($\overline{G}_{ww}$ from Eq. \ref{eq:Hyw_scale}) versus frequency for three cases are shown in Fig. \ref{fig:SAMM_eigenvalue_2sensors}. For SAMM-RR, the dominant eigenvalues do not change significantly for the different sensor combinations, which indicates the global nature of the flow field. We also note that the eigenvalues at the Rossiter frequencies are very close to those of SAMM-SPOD in Fig. \ref{fig:SAMM_SPOD_eigenvalues}.
	
\begin{figure*}[!htpb]
		\centering
		\begin{subfigure}[b]{1\textwidth}
			\centering
			\includegraphics[width=\textwidth]{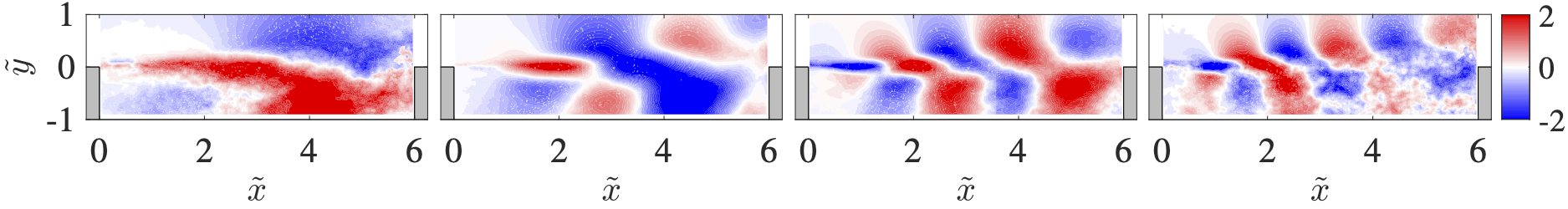}
			\vspace{-10pt}
			\caption{Mode shape of $u$, case 1}
			\vspace{0pt}
		\end{subfigure}
		\begin{subfigure}[b]{1\textwidth}
			\centering
			\includegraphics[width=\textwidth]{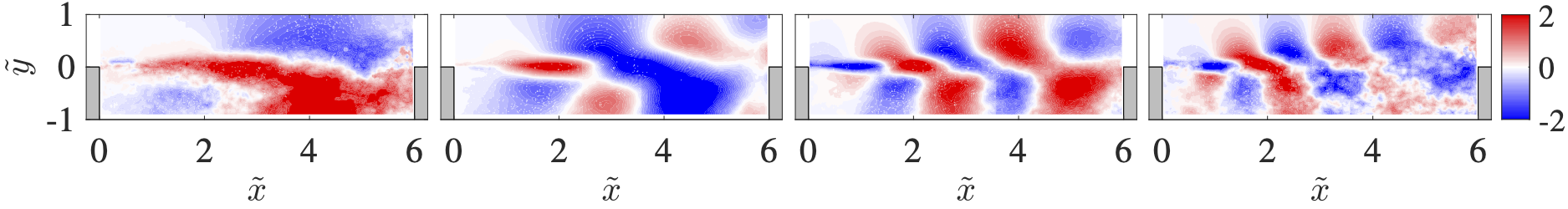}
			\vspace{-10pt}
			\caption{Mode shape of $u$, case 2}
		\end{subfigure}
		\begin{subfigure}[b]{1\textwidth}
			\centering
			\includegraphics[width=\textwidth]{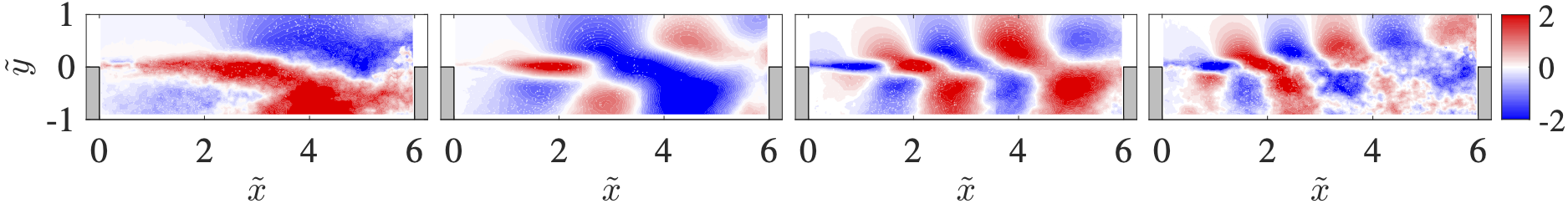}
			\vspace{-10pt}
			\caption{Mode shape of $u$, case 3}
		\end{subfigure}
		\begin{subfigure}[b]{1\textwidth}
			\centering
			\includegraphics[width=\textwidth]{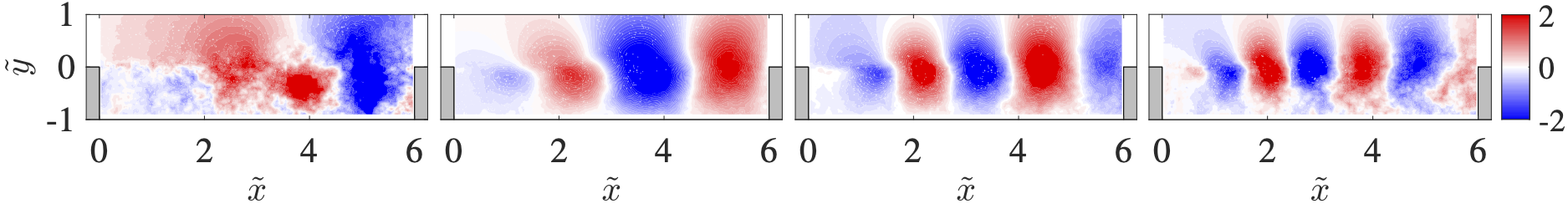}
			\vspace{-10pt}
			\caption{Mode shape of $v$, case 1}
			\vspace{0pt}
		\end{subfigure}
		\begin{subfigure}[b]{1\textwidth}
			\centering
			\includegraphics[width=\textwidth]{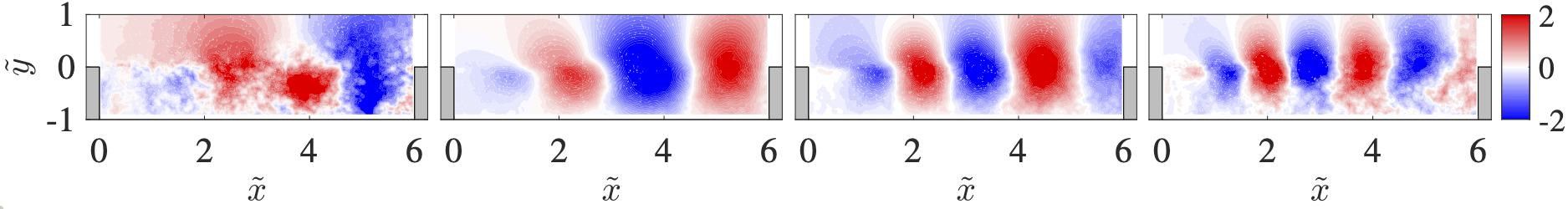}
			\vspace{-10pt}
			\caption{Mode shape of $v$, case 2}
		\end{subfigure}
		\begin{subfigure}[b]{1\textwidth}
			\centering
			\includegraphics[width=\textwidth]{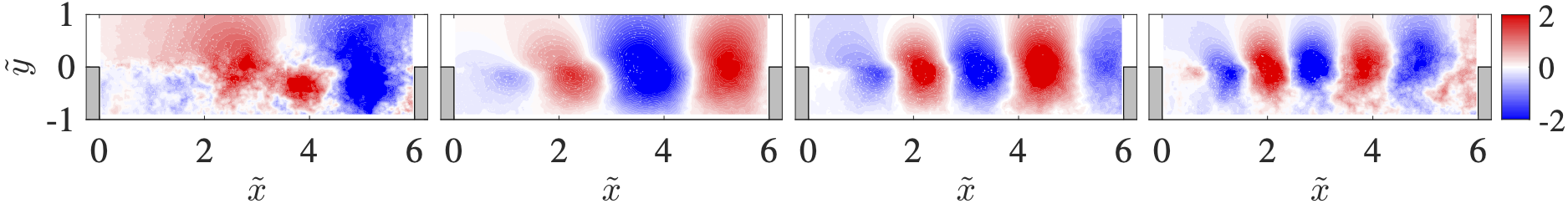}
			\vspace{-10pt}
			\caption{Mode shape of $v$, case 3}
		\end{subfigure}
		\caption{Rank-1 POD modes corresponding to first 4 Rossiter frequencies from SAMM-RR. Case 1: all sensors; case 2: $1^{st}$ and $10^{th}$ sensors; case 3: $1^{st}$ and $2^{nd}$ sensors.}
		\label{fig:SAMM_modes}
	\end{figure*}

	The dominant POD modes, located in the first column of $\overline{H}_{yw}$, associated with first four Rossiter frequencies are obtained. Again, we compare the results obtained for the 3 sensor combinations above in Fig. \ref{fig:SAMM_modes}. It is clear that these modes are very similar to the TR-PIV SPOD modes in Fig. \ref{fig:SPOD_modes}. Furthermore, the modes obtained by different combinations of sensors are almost identical. These observations indicate that the dynamics of the cavity flow is a global feature that can be captured by a sensor located anywhere inside the cavity (except at a pressure node). The good agreement between the SPOD modes and SAMMs modes shows that the SAMMs are capable of finding the dominant dynamic coherent flow structures without time resolved velocity data. 
	
	\subsection{Uncertainty}
	Here we assess the uncertainty of $G_{xy}$. While $G_{x_ix_j}$ can be accurately estimated using the standard periodogram method via the long time series of the TR pressure data, the cross correlations between unsteady TR pressure and NTR velocity must use Eq. \ref{eq:Rxy} followed by Eq. \ref{eq:GxyFFT}. The reduced number of terms in the ensemble average compared to that available with all TR data increases the random error. We can evaluate the impact of the number of averages qualitatively by examining the difference between using TR vs. NTR pressure data to estimate the correlations. The NTR pressure data sequence is obtained by retaining only those samples corresponding to the instances when the dual lasers pulse, thereby simulating the PIV data sequence. 
	
	\begin{figure}[!htpb]
		\centering
		\includegraphics[width=0.45\textwidth]{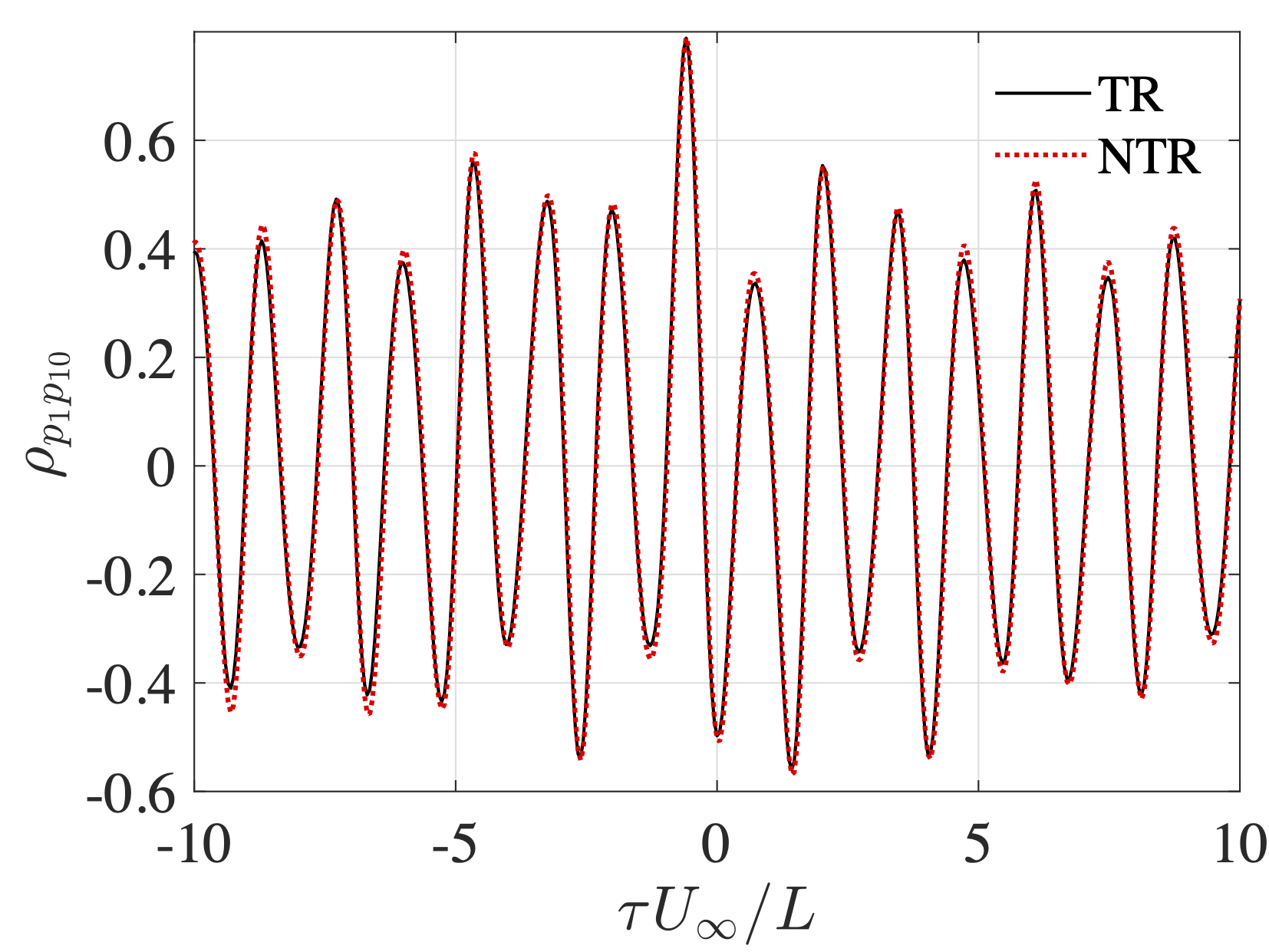}
		\caption{Zoomed-in comparison between the cross-correlation coefficient between pressure sensors 1 and 10 calculated using TR data and NTR data via Eq. \ref{eq:Rxy}.}
		\label{fig:R_narrow}
		\vspace{-0.0in}
	\end{figure}
	
	Fig. \ref{fig:R_narrow} shows a ``zoomed-in'' view of the cross correlation coefficient as a function of dimensionless time delay between pressure sensors 1 and 10 computed two ways.  The first, denoted as ``TR'', uses all TR pressure data, while the second, denoted as ``NTR'', uses a NTR pressure sequence combined with a second TR pressure data sequence. The one calculated using NTR data uses 4772 averages, corresponding to the number of PIV snapshots, while the TR data is computed using the entire time series of pressure data. A slight amplitude mismatch is noted at the local extrema, while the temporal locations of the extrema are accurately captured. Although the example shown is for the unsteady pressure signals, similar discrepancies are expected for the cross correlations between the unsteady pressure and the velocity field. 
	
\begin{figure}[!htpb]
		\centering
		\includegraphics[width=0.45\textwidth]{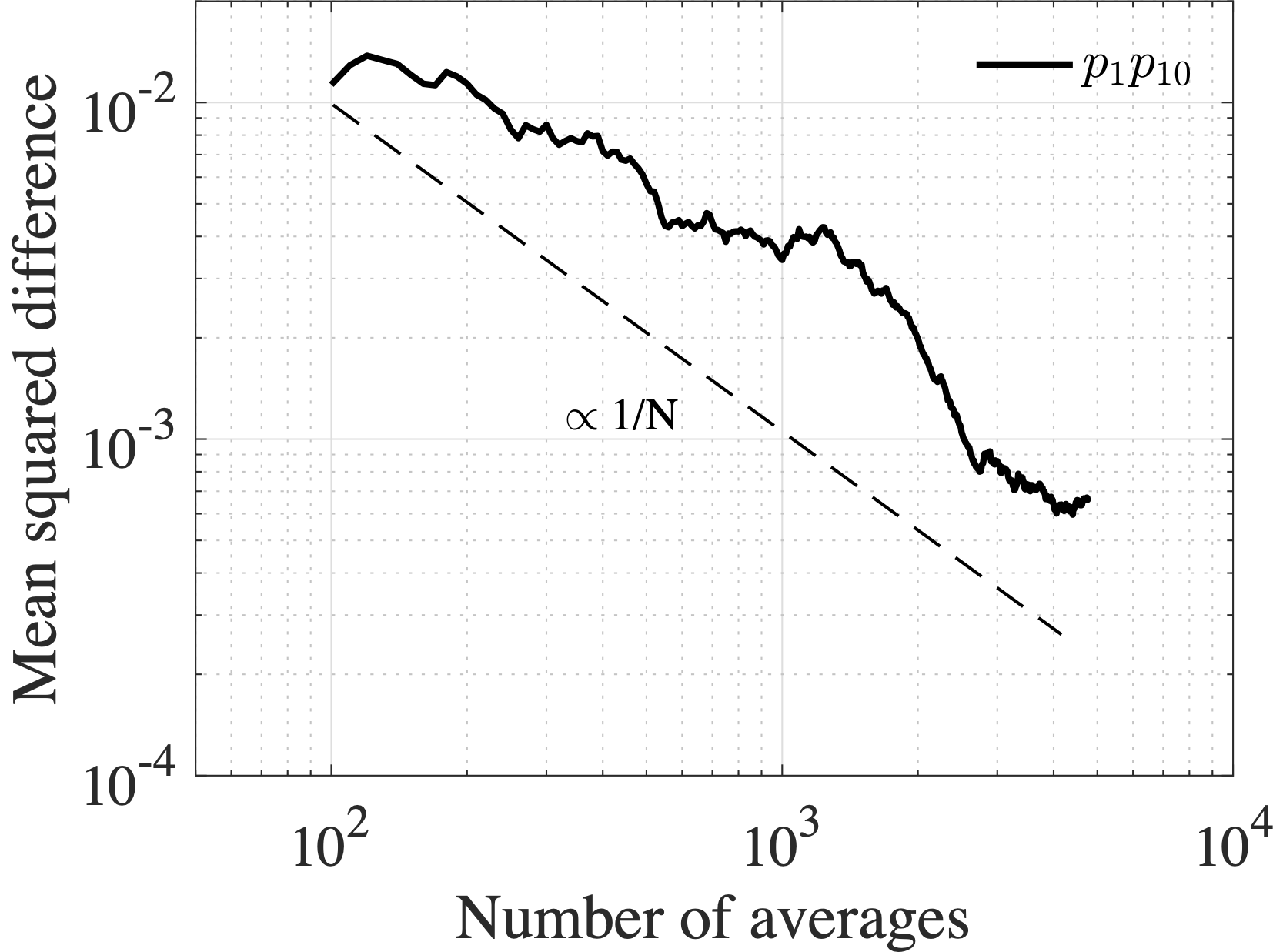}
		\caption{Mean-squared difference between the cross-correlation coefficient computed between pressure sensors 1 and 10 using NTR data versus TR data.
		}
		\label{fig:MSE}
		\vspace{-0.0in}
	\end{figure}

	\cite{BP} provide an estimate for the normalized rms random error in the cross correlation coefficient between two zero-mean TR signals, $x$ and $y$, under the assumption that both are band-limited white noise signals, as
	\begin{equation}
	\epsilon[\hat{R}_{xy}(\tau)]\approx\frac{1}{\sqrt{2BT_\text{total}}}[1+\rho_{xy}^{-2}(\tau)]^{0.5},
	\end{equation}
	where $B$ is the bandwidth, $T_\text{total}=N_\text{total}\Delta t$ is the total record length, and $\rho_{xy}$ is the dimensionless correlation coefficient. For TR data, the term $2BT_\text{total}\approx N_\text{total}$, so $\epsilon$ exhibits the expected inverse square root dependence on the number of averages. By increasing the number of PIV snapshots, the correlation functions calculated from the NTR data will converge to the ones calculated using TR data; thus, the random error can be reduced.  This is illustrated in Fig. \ref{fig:MSE}, where the mean-square difference between the two cross correlation coefficient estimates, which is expected to be close to the mean-squared error for an unbiased estimator, is $\propto 1/N_\text{averages}$.  Experimentation for the cavity flow case suggests that at least 1000 snapshots are required for statistical convergence.
	
	\begin{figure}[!htpb]
		\centering
		\includegraphics[width=0.45\textwidth]{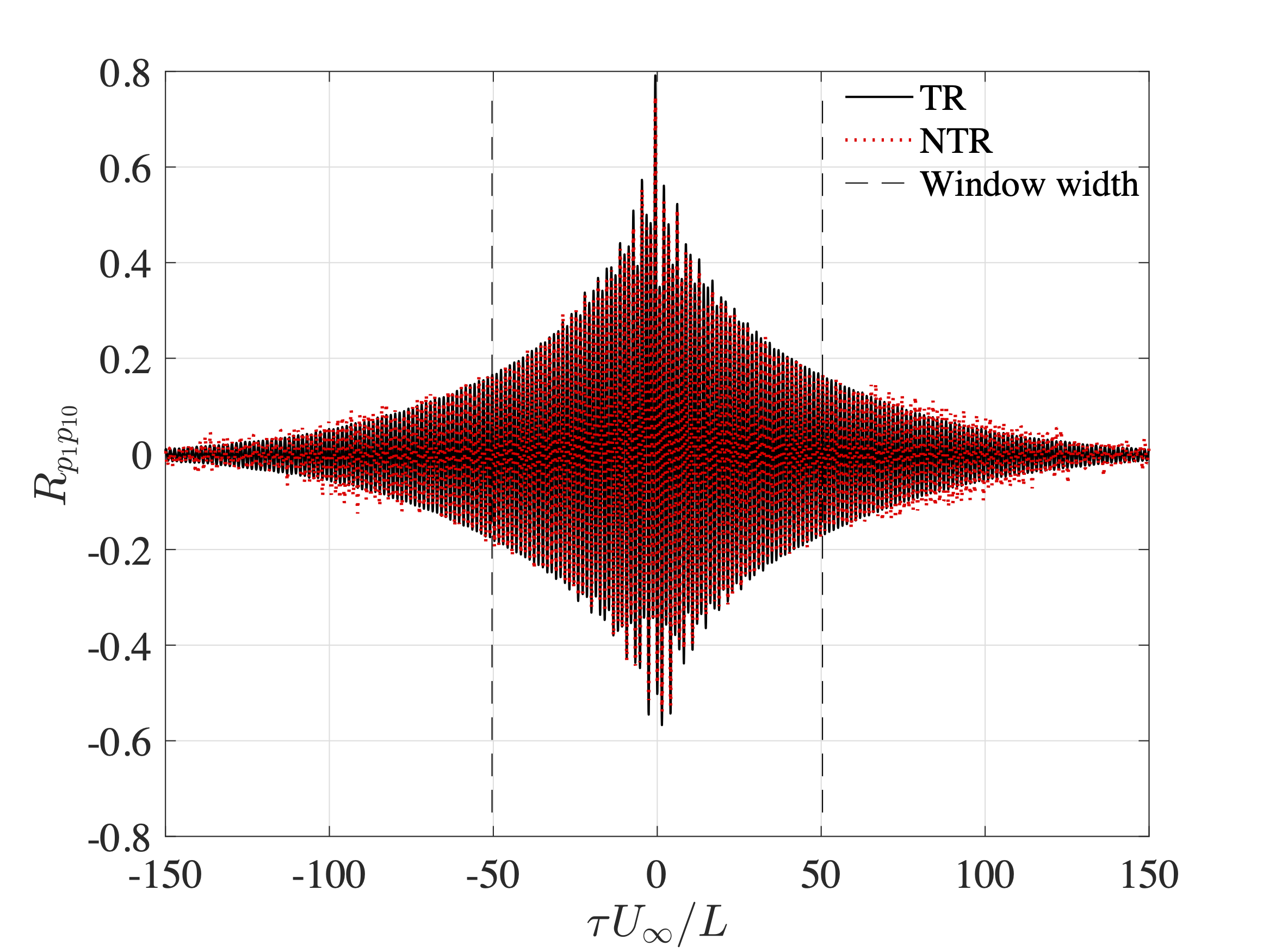}
		\caption{Zoomed-out comparison between the cross-correlation coefficient between pressure sensors 1 and 10 calculated using TR data and NTR data via Eq. \ref{eq:Rxy}.}
		\label{fig:Rp1p10_all}
	\end{figure}
	
	The other item to address with regards to uncertainty is the FFT of the windowed cross correlation.  A zoomed-out view of the cross-correlation in Fig. \ref{fig:R_narrow} is shown in Fig. \ref{fig:Rp1p10_all}. The magnitude of the correlation coefficient decreases for large $|\tau|$. However, it does not rapidly decay to zero due to the natural self-sustaining oscillations of the cavity flow. In practice, the maximum time delay, $\tau_\text{max}$, should be large enough so that the correlation coefficient decays to approximate noise levels and such that $1/(2\tau_\text{max})$ is the desired frequency resolution $f_s/N_{FFT}$.  For example, $N_{FFT}$ = 2048 and 4096 correspond to ${(\tau U_{\infty}/L)}_\text{max}$ = 50 and 100, respectively in Fig. \ref{fig:Rp1p10_all}, yielding correlation coefficients of $\approx$ 0.1 or less.  Noting the increasing discrepancy between for ${(\tau U_{\infty}/L)}_\text{max}>60$, we find $N_{FFT}$ = 2048 to be a suitable compromise.
	
	\section{Conclusions}
	In the current work, we present spectral analysis modal methods (SAMMs) to perform POD in both space and time using non-time-resolved Particle Image Velocity (PIV) data combined  with unsteady surface pressure measurements, and we illustrated their utility in flow-induced cavity oscillations at Mach 0.6. In this case, we used the TR unsteady surface pressure as the inputs and NTR velocity field from standard PIV as the outputs. To provide a comparison, we also obtained independent TR-PIV measurements at 16 kHz. Using a frequency-domain MIMO model combined with conditional spectral analysis, we can obtain independent inputs to the system and obtain a very good low-rank approximation of the dynamics obtained from frequency domain POD of the TR-PIV data. The eigenvector $A$ of the input auto/cross-spectra matrix is linked with the POD modes of the velocity fields through the transfer matrix $H$ between the pressure and velocity field. The mode shapes obtained from SPOD and the SAMMs are essentially the same. We also find that the modes can be captured very well with only 2 sensors from the cavity surface in SAMMs. The SAMMs outlined here can be used in other flows and laboratories where TR-PIV is not possible.  
	
	All results shown in this paper are computed using the open-source, SAMMs MATLAB implementation (uploaded to Matlabcentral/fileexchange).
	
	\begin{acknowledgements}
		This research was supported by the U.S. Air Force Office of Scientific Research (Grant
		FA9550-17-1-0380, Program Manager: Dr. Gregg Abate). The authors also gratefully acknowledge fruitful conversations with Profs. Peter Schmidt and Kunihiko Taira.
	\end{acknowledgements}
	
	
	\bibliographystyle{spbasic}      
	\bibliography{myrefs}   
	
	
\end{document}